# Generative AI to Generate Test Data Generators

Benoit Baudry, Khashayar Etemadi, Sen Fang, Yogya Gamage, Yi Liu, Yuxin Liu, Martin Monperrus, Javier Ron, André Silva, Deepika Tiwari

{baudry, khaes, senf, yogya, raphina, yuxinli, monperrus, javierro, andreans, deepikat}@kth.se

*KTH Royal Institute of Technology, Sweden*

## 1. Introduction

Software tests require data that is realistic, but not real. For example, banking applications cannot be tested with actual customer names and addresses. In these situations, developers rely on fake data generators, also known as fakers, to generate test data to be used in automated tests. Fakers exist in all programming languages. For example, the `faker` gem and `java-faker` are popular faking libraries for the Ruby and Java languages. Faking libraries usually include generators for names, phone numbers, and addresses. The development of test data generators is challenging, as they must consider several constraints. For example, name generators must capture the cultural sphere into which the system under test is being deployed. In many Spanish-speaking countries, a family name generator must output two names separated by a space. Another constraint relates to humor, as fakers have been proven to be a strong vector of healthy humor for bonding software development teams [1]. For an English-speaking developer, character names from *Star Trek* or *Seinfeld* are more exciting test data than John Doe, and there is support for this in faking libraries. Hence, the most advanced faking libraries contain data generators for specific languages, idioms, and cultures. These faking libraries are under constant evolution to stay in tune with testing constraints and the testing culture of the time.

Our intuition is that Large Language Models (LLMs) are powerful tools for supporting developers in generating high quality faking data. LLMs are unique systems that possibly encode: 1) domain expertise, 2) testing fluency, and 3) cultural literacy. Domain expertise is key in testing because most interesting data constraints come from the domain. For example, a French mobile phone number generator might output '08 790 60 001'. This would be incorrect, as a French number must start either with '06' or '07', and be split every two digits, e.g., '06 79 06 00 01'. For test fakers to engage developers, they should generate data that is both valid with respect to domain constraints and contains references to their language and culture. Finally, test data generators must be executable, and in some cases readily integratable into existing testing frameworks and their conventions. Our key intuition is that the generative power of LLMs can help master these three key aspects and be used for the generation of fake test data.

In this paper, *we study the original task of using LLMs for producing fake test data*. To the best of our knowledge, this promising area has never been studied.

We fully implement an approach based on the state-of-the-art LLM techniques for generating test data. To assess the feasibility of our approach, we curate real-world test data generation scenarios. For example, we use our approach to generate fake movie character names to help testing of a market-leading streaming service in China. We systematically assess the ability of the LLM to generate 1) test data that is fit for testing, and is culturally adequate; 2) executable code that synthesizes fake data; and 3) end-to-end code that is interoperable with state-of-the-art test data fakers.

To evaluate our approach, we have prompted the LLM 63 times to generate test data. The results indicate that LLMs are indeed able to generate fake test data that is realistic, compliant with data constraints, and readily usable in a testing context. When prompted for executable code and not only data, the LLM produces executable test data generators, ready to be used in test cases. To maximize ease of use and integration within test suites, it is important that the LLM has knowledge about existing faking frameworks: our experiments have also validated this aspect. In addition to these technical assessments of the generated data, we have also assessed the qualitative aspects of the generated test data. Our results indicate that LLMs are able to capture key cultural dimensions, including language and humor as part of the fake test data.

To sum up, our contributions are:

- an approach and prototype implementation to use LLMs for generating high-quality fake data.
- empirical evidence that LLMs are able to understand domain constraints such that the generated data is domain adequate, culturally fit and humorous.

> *Faking Libraries*
>
> The goal of a faking library is to generate realistic fake data, which is used as a substitute for real data within software tests. Fakers contain a rich collection of domain and locale-specific data, such as for the generation of user names or the generation of Chinese dishes. The first faker, an open-source library called `Data::Faker` introduced in 2005, produces fake data to test PERL programs. Its six generators provide data related to companies, dates and times, entities on the Internet such as email or IP addresses, Western names of persons, phone numbers, and US-specific street addresses. `Data::Faker` is designed to be flexible such that developers can extend it to define custom data generators. Over the years, multiple open-source faking libraries have emerged and are actively developed for all major programming languages, including Ruby, Python, Java, JavaScript, Rust, Haskell, and C++.
>
> In addition to conventional fakers, such as email generators, the developers of faking libraries incorporate data generators with strong cultural and humorous references [1]. When used within a test case, a quote from *Futurama* is likely as effective a string input as is Lorem ipsum text, with the added benefit that it is amusing to a developer who encounters it. Furthermore, good locale support within a faker can be helpful for developers who need test inputs in their native language, or to verify the internationalization of their system.

## 2. Test Data Generation with LLMs

Testing aims at exercising a software system realistically, without the system being deployed to an actual production environment. Instead of using production values in testing scenarios, developers rely on hard-coded data or fake data produced by so-called test fakers. In this work, we focus on generating test fakers, either in the form of pure data, or in the form of test modules that can be reused by developers to generate test data. In modern development, test fakers are typically provided as reusable faking libraries (see sidebar).

### 2.1. Overview

Figure 1 summarizes the key steps of our approach for generating fake test data with LLMs. First, we design prompts, which state the testing domain, the cultural constraints, as well as the programming language that the test generators should use. We now illustrate the concept with a realistic use case for our approach: testing a system for public administration. Such a system requires fake addresses that fulfill country-specific constraints, such as the language for street names or the specificities of postal codes. We propose three types of prompts, with different levels of complexity for the LLM-generated test generators, referred to as M1, M2, and M3. The M1 prompt asks the LLM to directly generate pure test data (e.g., addresses in Lisbon for a rental agency), with no code involved. M2 directs the LLM to generate a program that generates data (e.g., a Java program that generates addresses in Quimperlé for the French tax agency). With M3, we prompt the LLM to generate a program that generates fake data, and that aligns with a specific faking library (e.g., an address generator pluggable within `Faker.js`, to be used by a real estate company in Boston).

The second step is applicable to the outputs for M2 and M3. For both case, we execute the data generation code, as presented in Figure 1. Since the M2 and M3 prompts produce programs that generate data, it is necessary to actually run them to obtain the test data. Finally, the generated test data is used as input data within test cases for the system under test, such as the public administration software system.

The prompts can be done in English or in the local language of the system under test. Per our experiments, we recommend using the local language to maximize cultural adequacy, such as getting proper postal code or phone number formats. If the quality of the generated data is not satisfactory due to limited linguistic training resources, the English language can be employed as a fallback means of prompting.

### 2.2. M1: Directly Generate Test Data

In this mode, we use the ability of the LLM to generate pure test data. The outputs of M1 are directly used as inputs to test a software system. The core foundation of M1 is to craft a prompt that states: 1) the application domain of the system under test, 2) the expected natural language and cultural sphere, and, 3) the expected number of items.

For example, the prompt "生成十个中国武汉的假地址。" asks for fake addresses in Wuhan, China, that can be used as test data for the social security system for Wuhan residents. The expected outcome is a list of ten addresses that align with the Chinese address format and district names in Wuhan. Figure 1 presents an equivalent Portuguese prompt for addresses in Lisbon. A test harness takes the generated list of items and feeds it into the system under test.

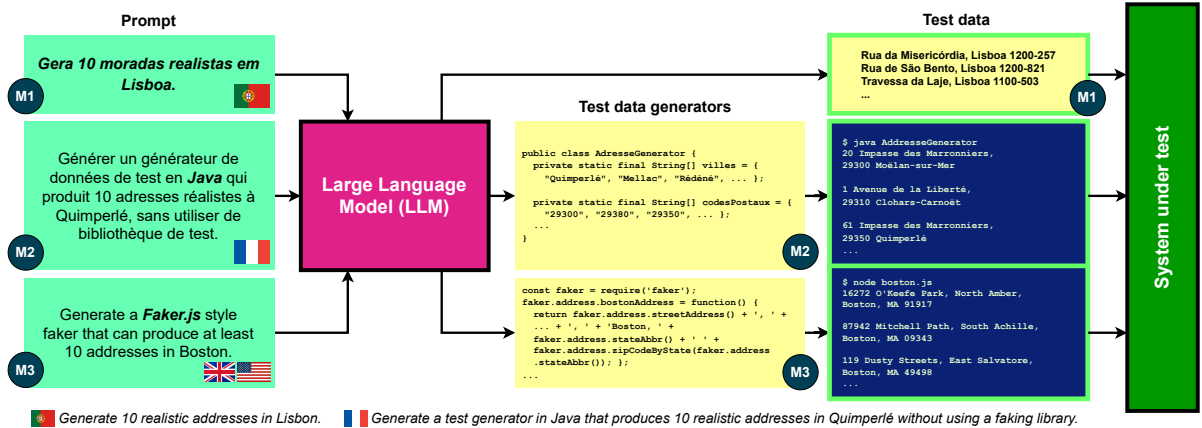

**FIGURE 1.** Overview of our approach for generating test data generators embedded in application domains and cultural spheres. We design three prompt types with the goal of generating test data. The prompts request for an output formatted as realistic fake data (M1), an automated data generator in a specific programming language (M2), or an automated data generator tailored to a specific faking library (M3). The output of the LLM is either fake data (M1), or a fake data generator (M2, M3). Our culturally diverse team of authors analyzes the adequacy of this output in order to evaluate the ability of the LLM to generate test data that is domain adequate and culturally adequate.

### 2.3. M2: Generate Executable Test Data Generators

Beyond raw test data generation, LLMs can also be employed to produce executable code, which can generate random fake testing data. We refer to this mode as M2. This executable code is then integrated by developers to generate fake data into their test suite. Figure 1 shows an example prompt for M2. The prompt has three main sections. First, a message guides the LLM to "synthesize a test data generator without using any library." Second, the prompt specifies the target programming language in which the data generator should be synthesized, such as "Java". The third component mentions the type and the cultural context of the data that should be generated, e.g., "the program should produce addresses in Quimperlé." M2 prompts leverage the assumed capability of LLMs to 1) understand the testing domain, and 2) generate complete, executable code [2].

### 2.4. M3: Generate Complete Interoperable Test Fakers

In the M3 mode, we aim at synthesizing end-to-end test data generators on top of existing faking libraries (see sidebar). The main motivation for this mode is to minimize the effort of integrating the test generators in an existing test suite. To that extent, M3 prompts are more effective than M2 prompts. This productivity boost happens thanks to the benefits of software reuse, here in the context of faking libraries.

Figure 1 shows an example prompt used for M3. The prompt asks the LLM to create a Javascript test data generator that specifically uses the `Faker.js` library. It also specifies the type of data that should be produced by the test data generator, such as "10 addresses in Boston".

## 3. Experimental Methodology

With the three modes of prompting described above, we generate test data and test data generators for various application domains. To this end, we draw upon the diverse backgrounds and expertise of co-authors, and guide the LLM to generate test data generators for applications in Chinese, Farsi, Portuguese, Sinhalese, French, Hindi, Spanish, and English. We select a frontier LLM, GPT-4, for our experiments and study its performance under the strictest conditions of a zero-shot setup. We devise three research questions to evaluate our novel approach of test data synthesis with LLMs:

- **RQ1:** To what extent is the LLM able to generate high quality, domain-adequate data?
- **RQ2:** To what extent is the LLM able to generate executable code that synthesizes fake data?
- **RQ3:** To what extent is the LLM able to generate end-to-end, interoperable test data fakers?

Our experimental artifacts can be found in https://github.com/ASSERT-KTH/lollm.

### 3.1. RQ1 Domain adequacy

In this RQ, we assess the ability of the LLM to generate high-quality test data that is appropriate for the specified application domain. In order to validate quality, we examine the outputs from the LLM, leveraging the cultural diversity of the authors originating from 3 continents, fluent in 7 mother tongues, and balanced over genders. This diversity allows us to match one or several authors' cultural backgrounds with the cultural context of the studied domain, in order to validate the cultural adequacy of the automatically generated test data. The matched experts check whether the synthesized fake data 1) is realistic, 2) is appropriate with respect to the semantics of the application domain, and 3) is consistent with the cultural dimensions specified in the prompt.

### 3.2. RQ2 Executability

In RQ2, we assess the ability of the LLM to synthesize executable code that generates fake data. To do that, we write M2 and M3 prompts and run the code produced by the LLM. We check whether this generated code can successfully be executed to completion. We also check for domain adequacy per the rule described in RQ1.

### 3.3. RQ3 Interoperability

For the final RQ, we evaluate the ability of the LLM to generate accurate, high-quality end-to-end test data fakers with the M3 mode. We start with the same evaluation criteria for adequacy and executability per RQ1 and RQ2. Additionally, we select one open-source project that uses a faking library in its test suite. We replace the original faker with the LLM-generated version. Finally, we run the full test suite of the project to verify that all the tests still pass with the generated fake data.

## 4. Experimental Results

We have prompted the LLM 63 times, in 8 different natural languages, and within 10 application domains. For the sake of brevity, we focus on a subset of domains and prompts, in each research question. The curious reader can browse our appendix repository for more fake test data generators.

### 4.1. RQ1: Data Adequacy

When prompting the LLM to generate pure test data, we discover a high level of cultural adequacy in 5 cases, and an absence of adequacy in 2 cases. We now discuss the cultural context and the adequacy of the LLM output for two domains.

*4.1.1. Case study: Adequacy of Chinese data for testing a streaming application.* Here, we are testing a streaming application, such as Netflix. We prompt GPT-4 to generate ten suitable names for the Chinese TV series *My Own Swordsman*, using the M1, M2, and M3 prompting modes. Next, three Chinese co-authors assess the generated names with respect to their cultural adequacy. According to our evaluation, all three modes of prompting can instruct GPT-4 to generate 10 fake names for *My Own Swordsman*. Specifically, the names generated by each prompt align with the background of the show and display full culture adequacy. For example, eight names from the M1 prompt are suitable for our TV show, such as 风流剑痴, 清风子, and 月影红. From our Chinese analysts' view, 风流剑痴, which means "A charming man who is passionate about swordsmanship" is considered the best one, while also being highly consistent with the mix of ancient culture and humor that characterizes the show.

Recently, several Chinese LLMs have been developed by the research divisions of companies such as Baidu and Alibaba. For further evaluation, we employ the same M1 prompt with two Chinese LLMs, ERNIE Bot and Qwen. We find that GPT-4 performs better than these two LLMs. Overall, although GPT-4 is not a Chinese LLM, it is the better choice for Chinese software testers if they want to obtain relevant fake data with respect to Chinese language and culture.

*4.1.2. Case study: Adequacy for low-resource languages.* In this case study, we are testing a travel application, such as TripAdvisor. We request GPT-4 to generate tourist attractions in Sri Lanka in Sinhalese, with M1 and M2 prompting modes. We observe that the generated results often include non-existent places. The following text presents a generated output where the locations are completely hallucinated: 'අරුගබේ', 'අරුගම්බේ කුමාර අවුස විහාරය', 'අරුගන්බේ කුමාර වෙනිකාව', 'දුක්බාගොඩ ලංචය'. We believe that the primary reason for this poor performance lies in the limited training of GPT-4 with Sinhalese text. To produce a more satisfactory output, the model would require training on a large volume of Sinhalese data, which is likely missing in the OpenAI training dataset. Overall, because of poor tokenization and lack of training data in Sinhalese, the generated data is of low quality. This case study highlights the limitation of our approach for low-resource languages. However, for all the other application domains with high-resource languages, we observe strong domain adequacy, including Chinese, French, Hindi, Portuguese, and Spanish.

> *Answer to RQ1*
>
> LLMs are able to generate high-quality test data. Our evaluation on 63 carefully designed case studies indicates that LLMs successfully capture the application domain of the test data as well as the cultural and linguistic constraints associated to it. This is good since software systems are designed and embedded in countries and cultures all over the world, while all being tested with the same rigor.

### 4.2. RQ2: Executability

We now focus on M2 and M3 prompts to evaluate the executability of the code generated by the LLM. We have performed 17 M2 and 25 M3 prompts, and in 29/42 cases, we obtain executable code. We now discuss two interesting case studies.

*4.2.1. Case study: Portuguese food and wine pairing.*
We aim to generate Ruby code that produces random, fake data, to test a food recommendation system such as Vivino. In this context, software developers expect the data to be realistic and correspond to culturally adequate suggestions. Ideally, the data constraints are explicit predicates in code that can be checked. We study the extent to which GPT-4 is able to generate an executable data constraint related to pairings between Portuguese food and wine, using the `faker` Ruby library (M3).

Listing 1 shows a snippet of the test data generator synthesized by GPT-4 that implements this data constraint. Within Portuguese dining, red wines are typically paired with meat, while white wines are paired with fish. On line 25 of the listing, the wine type is checked at runtime against the food type. The pair is kept if it complies with the wine-pairing constraint.

```ruby
require 'faker'

class PortuguesePairingGenerator
  PORTUGUESE_WINES = {
    'Vinho Verde' => :white,
    'Alvarinho' => :white,
    'Vinho Alentejo' => :red,
    'Porto' => :red,
    'Madeira' => [:red, :white]
  }

  PORTUGUESE_FOODS = {
    'Bacalhau a bras' => :fish,
    'Caldo verde' => :vegetable,
    'Francesinha' => :meat,
    'Arroz de Pato' => :meat,
    'Sardinhas assadas' => :fish
  }

  def self.generate
    wine, type = PORTUGUESE_WINES.to_a.sample
    food, food_type = PORTUGUESE_FOODS.to_a.sample

    # Pair white wine with fish and red wine with meat
    if (type == :white && food_type == :fish)
       || (type == :red && food_type == :meat)
      puts "Pair the #{wine} wine with #{food}."
    else
      generate
    end
  end
end

# Generate a pairing
PortuguesePairingGenerator.generate
```

**Listing 1.** Wine-pairing test data generator, generated by an LLM, with an embedded wine-pairing data constraint. In this example, we prompt GPT-4 to generate Ruby code with method M3: "(...) Please create a custom test data generator that generates wine-pairings between Portuguese wines and Portuguese foods."

This example shows that the capability of the model is two-fold: 1) it is aware of wine-pairing conventions, and 2) it is able to embed wine-pairing constraints in code. In total, we had success in generating executable `faker` code, and remarkably also found executable data constraints regarding food and wine types embedded in the generated code.

*4.2.2. Case study: Data Constraint in Farsi poetry - Testing applications using right-to-left scripts.* In this case study, we are testing a web publishing application that should support right-to-left scripts with constraints on the size of each line. For this, we employ the M2 mode to synthesize executable Java code that generates Farsi poetry in Masnavi style [3]. This type of poetry is written from right to left, and the lines of the poem should have approximately the same length. For this experiment, we use the following M2 prompt: "Generate a Java program without using any library that generates Farsi poem in Masnavi style as test data." The result is a Java application that successfully executes and generates two lines in Farsi. First, 'به یاد کسی میرویم - که در دل کسی میسویم', and second, 'به یاد آن روزگار - که با خود میبرد هموار'. The text is written in Farsi, which means it is right-to-left as expected. It also consists of lines with almost the same length. One limitation is that the generated text does not follow the rhythmic patterns of Farsi poetry, but we consider this constraint beyond the scope of the considered domain adequacy. Overall, this case study suggests that the LLM is able to generate executable code that produces proper Farsi text. The generated Farsi text can be useful for testing web applications displaying right-to-left text.

> *Answer to RQ2*
>
> We perform 42 prompts to assess the executability of test data generators produced by LLMs. The results indicate that LLMs are able to synthesize ready-to-use programs for generating test data. LLMs are able to reconcile the dual constraints of generating adequate test data in the considered domain, and generating source code that compiles and executes in a given programming language.

## 4.3. RQ3: Compatibility with Existing Faking Libraries

For this RQ, we prompt the LLM to extend an existing faking library, and integrate this extended library into the test suite of a real-world Java project. We target the test suite of a project called `sakai`, which is an open-source, feature-rich learning management system. `sakai` already uses the `java-faker` library in multiple test classes for generating fake names and placeholder text inputs. For example, lines 25 and 26 of Listing 2 show how the faker is used within the test class `ElasticSearchTest` to generate a fake name for a `Resource` object, such as `Jane Doe key keyboard`. Then, this object is used for testing the search implementation within the test case `testGetSearchSuggestions` (lines 30-36) to obtain search suggestions that contain the strings `key` and `keyboard` from an ElasticSearch service. The assertion on line 35 verifies that the suggestion list includes the recently created resource name, `Jane Doe key keyboard`.

We prompt the LLM in M3 mode to generate a `java-faker`-style generator that produces character names and quotes from the TV show *Merlin*. Per our expectations, the LLM generates code that follows the structure of the `java-faker` library, specifically a faker

```java
// LLM-generated extension of java-faker for Merlin
package com.github.javafaker;

public class Merlin {
  private final Faker faker;

  protected Merlin(Faker faker) {
    this.faker = faker;
  }

  public String character() {
    return faker.fakeValuesService().resolve(
      "merlin.characters", this, faker);
  }

  public String quote() {
    return faker.fakeValuesService().resolve(
      "merlin.quotes", this, faker);
  }
}
...........................................................

// Excerpt from the ElasticSearchTest class of sakai
public class ElasticSearchTest {
  String resourceName =
-   faker.name().name() + " key keyboard";
+   faker.merlin().character() + " key keyboard";

  ...
  @Test
  public void testGetSearchSuggestions() {
    String[] suggestions = elasticSearchService
    .getSearchSuggestions("keyboard", siteId, false);
    List suggestionList = Arrays.asList(suggestions);
    assertTrue(suggestionList.contains(resourceName));
  }
}
```

**Listing 2.** Lines 1-20: An extension of the `java-faker` library generated by the LLM to produce characters and quotes from the TV show *Merlin*. Lines 23-37: An excerpt from a test case in the project `sakai` which uses the `java-faker` library to generate fake resource names. We replace the existing call to generate fake names (line 26) with names from characters in *Merlin* (line 27).

class called `Merlin.java` (lines 1 to 20 in Listing 2), and a `merlin.yml` file containing character names and quotes. Moreover, the two generated files follow the same pattern as the existing generators within `java-faker`. Next, we extend `java-faker` with these two new files, and replace the existing `java-faker` version in `sakai` with this extended version. We update the test class `ElasticSearchTest` to generate fake names from characters in *Merlin*, as illustrated on lines 26 and 27 of Listing 2. Finally, we compile the project and execute the test suite, which now uses this extended faker.

Within the class `ElasticSearchTest`, 3 test cases call the LLM-generated faker, and 5 assertions assess behavior using this fake data. We observe that the complete end-to-end integration works seamlessly: 1) the test suite compiles and runs, and 2) all the test cases successfully pass with the extended `java-faker` library. From a testing perspective, using a resource called `Uther Pendragon` from the generated *Merlin* faker is as effective as using a conventional *"Jane Doe"* resource. This is strong evidence that the LLM is capable of successfully generating fakers that are ready to be used by developers, while engaging them even more with their tasks.

> *Answer to RQ3*
>
> LLMs encode knowledge about popular faking libraries, used to test thousands of software projects. This knowledge can be leveraged to generate new fakers, directly interoperable with test suites. To the best of our knowledge, our paper is the first to bridge the creative power of LLMs and the hard engineering constraints of data faking.

## 5. Assorted Remarks

In this section, we summarize the important insights from our experiments, which may fuel further research in the application of generative AI for fake test data generation.

### 5.1. LLM Selection

Our approach is designed to work with off-the-shelf LLMs. Thanks to instruction tuning, LLMs are able to follow natural language instructions, which, in our case, include the description of an application domain as well as essential cultural framing for the test data generation. Our findings indicate that general-purpose LLMs, such as GPT-4, are suitable for this task, and we believe that open-source LLMs would also demonstrate good capabilities. In a low-resource language or cultural settings, the choice of the LLM may result in higher performance variability, as evidenced by our experiments with Sinhalese subsubsection 4.1.2.

### 5.2. LLM Randomness

The code and data generated by LLMs are not deterministic, with randomness increasing when a high temperature is used. This is natural non-determinism in the test data that is produced. Therefore, as long as the generated test data is valid, the randomness in LLMs' output is a strength of our approach. As discussed in section 4, our 63 experiments show the general validity of the test data generated. Overall, we recommend embracing the randomness of LLMs to generate diverse test data, while also ensuring the generated data complies with required constraints with appropriate techniques.

### 5.3. Harness LLMs in test suites

In case study 1 (subsubsection 4.1.1), we observed that the test data produced using the M1-level prompt is of better quality compared to that generated by M2 and M3-level prompts. This disparity might be attributed to the increased complexity imposed by the M2 and M3-level prompts, potentially complicating the task for LLMs. This suggests an innovative approach: instead of employing LLMs to create a generator for generating test data, LLMs themselves could be directly utilized for generating test data in various test scenarios. This approach simplifies the process, increasing the validity of the generated data and making a responsible usage of generative AI for test data generation.

## 6. Related Work

Large Language Models have found application in various phases of software engineering, from the generation of specifications to the maintenance of legacy software [2]. Yet there are caveats to the use of LLMs in software engineering tasks, owing to their unpredictability, and issues such as potential data leakage [4]. Ouyang et al. [5] highlight that the non-determinism of LLMs can negatively impact code generation, producing semantically and syntactically different, potentially incorrect, code based on hyper-parameter configurations. In the same vein, Poesia et al. [6] propose an approach to enforce constraints on the code generated by LLMs, including syntax and variable typing. In this paper, we leverage the creative power of LLMs to synthesize wide-ranging test data, and their programming power to produce executable test data generators.

Several studies have experimented with LLMs and machine learning models as tools to aid software testing, including the automated generation of unit tests [7]. MockSniffer [8] uses machine learning to recommend components that may be substituted by mocks within unit tests. QTypist [9] uses LLMs to generate context-aware text inputs for testing mobile application interfaces. Tan and colleagues explore the use of Recurrent Neural Networks (RNNs) [10] and LLMs [11] for generating synthetic, representative test data for the Norwegian national population registry. Generative Adversarial Networks (GANs) have been used to anonymize test data used in the healthcare domain [12]. Similar to these studies, we utilize LLMs to automatically produce realistic synthetic test data. A core novelty of our work is the use of LLMs in the context of fakers, to automatically generate executable code that produces test data.

Many researchers have explored the cultural (in)adequacies exhibited by LLM outputs. Cao and colleagues [13] discover that ChatGPT performs poorly in non-American contexts. Naous et al. [14] analyze the cultural adaptability of LLMs, concluding that Arabic LLMs default to Western cultures. Chen et al. [15] report on the inadequate performance of several LLMs in understanding Chinese humor, including the detection of punchlines. In this work, we have explored diverse dimensions of cultural adequacy within test data. While the LLM performs very well at data generation, we observe that cultural adequacy varies depending on the task and the language used for prompting.

## 7. Conclusion

In this paper, we have addressed the original problem statement of generating test data generators with LLMs. To validate this far-reaching idea, we have performed an in-depth study into the capabilities of LLMs for generating test data, with attention to both hard software testing requirements and soft cultural requirements such as cultural adequacy. Our experimental results clearly indicate that current LLMs are able to succeed in this task. Over 63 prompts, we successfully obtain a large majority of good data generators, that can execute and produce valuable test data. As a final proof-of-concept, we integrate an LLM-generated data faker in the test suite of a mature and well-tested Java project. The complete success of this proof-of-concept indicates that LLM-generated test fakers can support serious and engaging software testing. Overall, our research opens a promising avenue for the use of generative models for generating data that is both adequate for testing and is culturally relevant. As future work, we envision using other types of prompting techniques, such as few-shot and chain-of-thought strategies, to improve the capability of the test data fakers generated by LLMs. It would also be interesting to compare the quality, diversity, and cost of LLM-generated fake data against manually-crafted fakers.

**Benoit Baudry** is a cheese-aficionado, humorous, and artistic Professor in Software Technology at the Université de Montréal. His research focuses on automated software engineering, software diversity and software testing. He favors exploring code execution over code on disk. He received his PhD in 2003 from the University of Rennes, France. He was a research scientist at INRIA (France) from 2004 to 2017, and a Professor at the KTH Royal Institute of Technology (Sweden) from 2017 to 2023. Contact him at benoit.baudry@umontreal.ca.

**Khashayar Etemadi** is a Ph.D. student at KTH Royal Institute of Technology, Sweden. He received his MSc and BSc degrees in Software Engineering from Sharif University of Technology, Iran. His research is focused on explainable software bots, ML4SE, and program analysis. Contact him at khaes@kth.se.

**Sen Fang** received the master degree in electronics and communication engineering from Central China Normal University in 2020. His research interests lie in the intersection of software engineering and machine learning, particularly LLMs for code. Contact him at sfang9@ncsu.edu.

**Yogya Gamage** is a research engineer at KTH Royal Institute of Technology, Stockholm, Sweden. Her current research focuses on software hardening, software supply chain security, and program repair. She received a bachelor's degree in computer science and engineering from the University of Moratuwa, Sri Lanka. Contact her at yogya.gamage@umontreal.ca.

**Yi Liu** is a research assistant at KTH Royal Institute of Technology, Stockholm, Sweden. Her research interests include the intersection between blockchain and security. She received a bachelor's degree from Sichuan University and is now pursuing a master's degree at Stockholm University. Contact her at raphina@kth.se.

**Yuxin Liu** is a Ph.D. student at KTH Royal Institute of Technology, Stockholm, Sweden. Her research interests include software engineering, software analysis, and software package management. Yuxin received her MSc in software engineering from Harbin Institute of Technology. Contact her at yuxinli@kth.se.

**Martin Monperrus** is a prolific award-winning Professor of Software Technology and 2024 PhD Supervisor of the Year at KTH Royal Institute of Technology, Sweden. His research lies in the field of software engineering with a current focus on automatic program repair, AI on code and program hardening. He received a Ph.D. from the University of Rennes, and a Master's degree from Compiègne University of Technology. Homepage: https://www.monperrus.net/martin/

**Javier Ron** is a Ph.D. student at KTH Royal Institute of Technology. He received his BSc degree from ESPOL University in Ecuador and his MSc degree from KTH Royal Institute of Technology. His research interest lies in the intersection of software engineering, distributed systems and game development. Contact him at javierro@kth.se.

**André Silva** is a Ph.D. student at KTH Royal Institute of Technology, Stockholm, 11428, Sweden. His research interests include the intersection of automatic program repair and machine learning. Silva received his M.Sc. in Computer Science from Instituto Superior Técnico, Universidade de Lisboa, Lisbon, Portugal. Contact him at andreans@kth.se.

**Deepika Tiwari** is a Ph.D. student at KTH Royal In- stitute of Technology, working on software testing. Her research focuses on automatic software test generation, production monitoring, and software humor. Contact her at deepikat@kth.se.